\input harvmac
\noblackbox
\newcount\figno
\figno=0
\def\fig#1#2#3{
\par\begingroup\parindent=0pt\leftskip=1cm\rightskip=1cm\parindent=0pt
\baselineskip=11pt \global\advance\figno by 1 \midinsert
\epsfxsize=#3 \centerline{\epsfbox{#2}} \vskip 12pt {\bf Fig.
\the\figno:} #1\par
\endinsert\endgroup\par
}
\def\figlabel#1{\xdef#1{\the\figno}}
\def\encadremath#1{\vbox{\hrule\hbox{\vrule\kern8pt\vbox{\kern8pt
\hbox{$\displaystyle #1$}\kern8pt} \kern8pt\vrule}\hrule}}

\input epsf

\overfullrule=0pt

%
\def\tilde{\widetilde}
\def\bar{\overline}

%
\def\inbar{\,\vrule height1.5ex width.4pt depth0pt}
\def\IB{\relax{\rm I\kern-.18em B}}
\def\IC{\relax\hbox{$\inbar\kern-.3em{\rm C}$}}
\def\ID{\relax{\rm I\kern-.18em D}}
\def\IE{\relax{\rm I\kern-.18em E}}
\def\IF{\relax{\rm I\kern-.18em F}}
\def\IG{\relax\hbox{$\inbar\kern-.3em{\rm G}$}}
\def\IH{\relax{\rm I\kern-.18em H}}
\def\II{\relax{\rm I\kern-.18em I}}
\def\IK{\relax{\rm I\kern-.18em K}}
\def\IL{\relax{\rm I\kern-.18em L}}
\def\IM{\relax{\rm I\kern-.18em M}}
\def\IN{\relax{\rm I\kern-.18em N}}
\def\IO{\relax\hbox{$\inbar\kern-.3em{\rm O}$}}
\def\IP{\relax{\rm I\kern-.18em P}}
\def\IQ{\relax\hbox{$\inbar\kern-.3em{\rm Q}$}}
\def\IR{\relax{\rm I\kern-.18em R}}
\font\cmss=cmss10 \font\cmsss=cmss10 at 7pt
\def\IZ{\relax\ifmmode\mathchoice
{\hbox{\cmss Z\kern-.4em Z}}{\hbox{\cmss Z\kern-.4em Z}}
{\lower.9pt\hbox{\cmsss Z\kern-.4em Z}} {\lower1.2pt\hbox{\cmsss
Z\kern-.4em Z}}\else{\cmss Z\kern-.4em Z}\fi}
\def\IGa{\relax\hbox{${\rm I}\kern-.18em\Gamma$}}
\def\IPi{\relax\hbox{${\rm I}\kern-.18em\Pi$}}
\def\ITh{\relax\hbox{$\inbar\kern-.3em\Theta$}}
\def\IOm{\relax\hbox{$\inbar\kern-3.00pt\Omega$}}

\def\p{\partial}

\font\zfont = cmss10 
\font\litfont = cmr6 
\def\bigone{\hbox{1\kern -.23em {\rm l}}}
\def\ZZ{\hbox{\zfont Z\kern-.4emZ}}

\def\half{{\litfont {1 \over 2}}}

\def\CM{{\cal M}}

\def\a{\alpha}
\def\b{\beta}
\def\d{\delta}

\def\vt{\vartheta}

\def\l{\lambda}
\def\m{\mu}

\def\s{\sigma}

\def\u{\upsilon}
\def\p{\phi}

\def\S{\Sigma}

\def\ch{{\rm ch}}
\def\IR{\relax{\rm I\kern-.18em R}}
\def\I1{\relax{\rm I\kern-.6em 1}}
\def\Dsl{\,\raise.15ex\hbox{/}\mkern-13.5mu D}
\def\Gsl{\,\raise.15ex\hbox{/}\mkern-13.5mu G}
\def\Csl{\,\raise.15ex\hbox{/}\mkern-13.5mu C}
\font\cmss=cmss10 \font\cmsss=cmss10 at 7pt
\def\pa{\partial}
\def\pam{\partial_-}

\def\cp{{\cal P}}

\def\ch{{\cal H}}
\def\cc{{\cal C}}
\def\cm{{\cal M}}
\def\cn{{\cal N}}
\def\ti{\tilde{I}}

\font\zfont = cmss10 
\font\litfont = cmr6 
\def\bigone{\hbox{1\kern -.23em {\rm l}}}
\def\ZZ{\hbox{\zfont Z\kern-.4emZ}}
\def\half{{\litfont {1 \over 2}}}

\def\CM{{\cal M}}

%
%
%
\message{S-Tables Macro v1.0, ACS, TAMU (RANHELP@VENUS.TAMU.EDU)}
%
%
\newhelp\stablestylehelp{You must choose a style between 0 and 3.}%
\newhelp\stablelinehelp{You should not use special hrules when stretching
a table.}%
\newhelp\stablesmultiplehelp{You have tried to place an S-Table inside
another
S-Table.  I would recommend not going on.}%
%
%
\newdimen\stablesthinline
\stablesthinline=0.4pt
\newdimen\stablesthickline
\stablesthickline=1pt
%
%
\newif\ifstablesborderthin
\stablesborderthinfalse
\newif\ifstablesinternalthin
\stablesinternalthintrue
\newif\ifstablesomit
\newif\ifstablemode
\newif\ifstablesright
\stablesrightfalse
%
%
\newdimen\stablesbaselineskip
\newdimen\stableslineskip
\newdimen\stableslineskiplimit
%
%
\newcount\stablesmode
\newcount\stableslines
\newcount\stablestemp
\stablestemp=3
\newcount\stablescount
\stablescount=0
\newcount\stableslinet
\stableslinet=0
%
%
%
\newcount\stablestyle
\stablestyle=0
%
%
\def\stablesleft{\quad\hfil}%
\def\stablesright{\hfil\quad}%
%
%
\catcode`\|=\active%
%
%
\newcount\stablestrutsize
\newbox\stablestrutbox
\setbox\stablestrutbox=\hbox{\vrule height10pt depth5pt width0pt}
\def\stablestrut{\relax\ifmmode%
                         \copy\stablestrutbox%
                       \else%
                         \unhcopy\stablestrutbox%
                       \fi}%
%
%
\newdimen\stablesborderwidth
\newdimen\stablesinternalwidth
\newdimen\stablesdummy
\newcount\stablesdummyc
\newif\ifstablesin
\stablesinfalse
%
%
\def\begintable{\stablestart%
  \stablemodetrue%
  \stablesadj%
  \halign%
  \stablesdef}%
\def\stablesadj{%
  \ifcase\stablestyle%
    \hbox to \hsize\bgroup\hss\vbox\bgroup%
  \or%
    \hbox to \hsize\bgroup\vbox\bgroup%
  \or%
    \hbox to \hsize\bgroup\hss\vbox\bgroup%
  \or%
    \hbox\bgroup\vbox\bgroup%
  \else%
    \errhelp=\stablestylehelp%
    \errmessage{Invalid style selected, using default}%
    \hbox to \hsize\bgroup\hss\vbox\bgroup%
  \fi}%
\def\stablesend{\egroup%
  \ifcase\stablestyle%
    \hss\egroup%
  \or%
    \hss\egroup%
  \or%
    \egroup%
  \or%
    \egroup%
  \else%
    \hss\egroup%
  \fi}%
\def\stablestart{%
  \ifstablesin%
    \errhelp=\stablesmultiplehelp%
    \errmessage{An S-Table cannot be placed within an S-Table!}%
  \fi
  \global\stablesintrue%
  \global\advance\stablescount by 1%
  \message{<S-Tables Generating Table \number\stablescount}%
  \begingroup%
  \stablestrutsize=\ht\stablestrutbox%
  \advance\stablestrutsize by \dp\stablestrutbox%
  \ifstablesborderthin%
    \stablesborderwidth=\stablesthinline%
  \else%
    \stablesborderwidth=\stablesthickline%
  \fi%
  \ifstablesinternalthin%
    \stablesinternalwidth=\stablesthinline%
  \else%
    \stablesinternalwidth=\stablesthickline%
  \fi%
  \tabskip=0pt%
  \stablesbaselineskip=\baselineskip%
  \stableslineskip=\lineskip%
  \stableslineskiplimit=\lineskiplimit%
  \offinterlineskip%
  \def\borderrule{\vrule width \stablesborderwidth}%
  \def\internalrule{\vrule width \stablesinternalwidth}%
  \def\thinline{\noalign{\hrule height \stablesthinline}}%
  \def\thickline{\noalign{\hrule height \stablesthickline}}%
  \def\trule{\omit\leaders\hrule height \stablesthinline\hfill}%
  \def\ttrule{\omit\leaders\hrule height \stablesthickline\hfill}%
  \def\tttrule##1{\omit\leaders\hrule height ##1\hfill}%
  \def\stablesel{&\omit\global\stablesmode=0%
    \global\advance\stableslines by 1\borderrule\hfil\cr}%
  \def\el{\stablesel&}%
  \def\elt{\stablesel\thinline&}%
  \def\eltt{\stablesel\thickline&}%
  \def\elttt##1{\stablesel\noalign{\hrule height ##1}&}%
  \def\elspec{&\omit\hfil\borderrule\cr\omit\borderrule&%
              \ifstablemode%
              \else%
                \errhelp=\stablelinehelp%
                \errmessage{Special ruling will not display properly}%
              \fi}%
  \def\stmultispan##1{\mscount=##1 \loop\ifnum\mscount>3 \stspan\repeat}%
  \def\stspan{\span\omit \advance\mscount by -1}%
  \def\multicolumn##1{\omit\multiply\stablestemp by ##1%
     \stmultispan{\stablestemp}%
     \advance\stablesmode by ##1%
     \advance\stablesmode by -1%
     \stablestemp=3}%
  \def\multirow##1{\stablesdummyc=##1\parindent=0pt\setbox0\hbox\bgroup%
    \aftergroup\emultirow\let\temp=}
  \def\emultirow{\setbox1\vbox to\stablesdummyc\stablestrutsize%
    {\hsize\wd0\vfil\box0\vfil}%
    \ht1=\ht\stablestrutbox%
    \dp1=\dp\stablestrutbox%
    \box1}%
  \def\stpar##1{\vtop\bgroup\hsize ##1%
     \baselineskip=\stablesbaselineskip%
     \lineskip=\stableslineskip%

\lineskiplimit=\stableslineskiplimit\bgroup\aftergroup\estpar\let\temp=}%
  \def\estpar{\vskip 6pt\egroup}%
  \def\stparrow##1##2{\stablesdummy=##2%
     \setbox0=\vtop to ##1\stablestrutsize\bgroup%
     \hsize\stablesdummy%
     \baselineskip=\stablesbaselineskip%
     \lineskip=\stableslineskip%
     \lineskiplimit=\stableslineskiplimit%
     \bgroup\vfil\aftergroup\estparrow%
     \let\temp=}%
  \def\estparrow{\vfil\egroup%
     \ht0=\ht\stablestrutbox%
     \dp0=\dp\stablestrutbox%
     \wd0=\stablesdummy%
     \box0}%
  \def|{\global\advance\stablesmode by 1&&&}%
  \def\|{\global\advance\stablesmode by 1&\omit\vrule width 0pt%
         \hfil&&}%
  \def\vt{\global\advance\stablesmode by 1&\omit\vrule width
\stablesthinline%
          \hfil&&}%
  \def\vtt{\global\advance\stablesmode by 1&\omit\vrule width
\stablesthickline%
          \hfil&&}%
  \def\vttt##1{\global\advance\stablesmode by 1&\omit\vrule width ##1%
          \hfil&&}%
  \def\vtr{\global\advance\stablesmode by 1&\omit\hfil\vrule width%
           \stablesthinline&&}%
  \def\vttr{\global\advance\stablesmode by 1&\omit\hfil\vrule width%
            \stablesthickline&&}%
  \def\vtttr##1{\global\advance\stablesmode by 1&\omit\hfil\vrule width
##1&&}%
  \stableslines=0%
  \stablesomitfalse}
\def\stablesdef{\bgroup\stablestrut\borderrule##\tabskip=0pt plus 1fil%
  &\stablesleft##\stablesright%
  &##\ifstablesright\hfill\fi\internalrule\ifstablesright\else\hfill\fi%
  \tabskip 0pt&&##\hfil\tabskip=0pt plus 1fil%
  &\stablesleft##\stablesright%
  &##\ifstablesright\hfill\fi\internalrule\ifstablesright\else\hfill\fi%
  \tabskip=0pt\cr%
  \ifstablesborderthin%
    \thinline%
  \else%
    \thickline%
  \fi&%
}%
\def\endtable{\advance\stableslines by 1\advance\stablesmode by 1%
   \message{- Rows: \number\stableslines, Columns:  \number\stablesmode>}%
   \stablesel%
   \ifstablesborderthin%
     \thinline%
   \else%
     \thickline%
   \fi%
   \egroup\stablesend%
\endgroup%
\global\stablesinfalse}
%

\lref\dm{D.R. Morrison, ``The Geometry underlying mirror
symmetry," alg-geom/9608006.}
\lref\grha{P.~ Griffiths and J.~ Harris, {\it Principles of
Algebraic Geometry,}  J.Wiley and Sons, 1978. }
\lref\hw{C. Hull and E. Witten, ``Supersymmetric sigma models and
the heterotic string,'' Phys. Lett. {\bf 160B}  (1985), 398.}
\lref\hub{T. H\"{u}bsch, {\it Calabi Yau Manifolds,} World
Scientific, 1992}
\lref\bbs{K. Becker, M. Becker and A. Strominger, ``Fivebranes,
Membranes and Non-Perturbative String Theory,''  Nucl. Phys. {\bf
B456} (1995) 130, hep-th/9507158.}
\lref\msw{J. Maldacena, A. Strominger and E. Witten, ``Black Hole
Entropy in M Theory,'' J. High Energy Phys. {\bf 12} (1997) 002,
hep-th/9711053.}
\lref\cfg{S. Cecotti, S. Ferrara  and L. Girardello, ``Geometry of
Type-II Superstrings and the Moduli of Superconformal Field
Theories,''  Int. J. Mod. Phys. {\bf A4}(1989) 2475.}
\lref\wittfive{E. Witten, ``Five-Brane Effective Action In
M-Theory,'' J. Geom. Phys. {\bf 22} (1997) 103; hep-th/9610234.}
\lref\hmm{J.A. Harvey, R. Minasian and G. Moore, ``Non-abelian
Tensor-multiplet Anomalies,'' J. High Energy Phys. {\bf 09} (1998)
004; hep-th/9808060.}
\lref\mcl{R.C. McLean, ``Deformations of calibrated
submanifolds,'' Comm. Anal. Geom. {\bf 6} (1998) 705.}
\lref\wof{E. Witten, ``On Flux Quantization in M-Theory and the
Effective Action," J. Geom. Phys. {\bf 22} (1997) 1;
hep-th/9609122. }
\lref\hita{N.J. Hitchin, ``The moduli space of special Lagrangian
submanifolds,'' math.DG/9711002.}
\lref\hitb{N.J. Hitchin, ``The moduli space of complex Lagrangian
submanifolds,'' math.DG/9901069.}
\lref\hl{F.R.~Harvey and H.B.~Lawson, ``Calibrated geometries,''
Acta Math.{\bf 148} (1982), 47. }
\lref\harv{F.R. Harvey, {\it Spinors and calibrations,} Academic
Press, 1990.}
\lref\gross{M. Gross, ``Special Lagrangian Fibrations II:
Geometry,'' math.AG/9809072.}
\lref\lm{H.B.~Lawson and M.L.~Michelsohn, {\it Spin geometry,}
 Princeton University Press, 1989.}
\lref\gipa{G.W.~Gibbons and G.~Papadopoulos, ``Calibrations and
intersecting branes,'' hep-th/9803163.}
\lref\fof{J.M. Figueroa-O'Farrill, ``Intersecting brane
geometries,'' hep-th/9806040.}
\lref\bmoy{K. Becker, M.Becker, D. Morrison, H. Ooguri, Y. Oz and
Z. Yin, ``Supersymmetric Cycles in Exceptional Holonomy Manifolds
and Calabi-Yau 4-Folds,'' Nucl. Phys. {\bf B480} (1996) 225;
hep-th/9608116.}
\lref\mmt{R. Minasian, G. Moore and D. Tsimpis, ``Calabi-Yau black
holes and (0,4) sigma models'' hep-th/9904217.}
\lref\dlm{M. J. Duff, J.T. Liu and R. Minasian, ``Eleven
Dimensional Origin of String/String Duality: A One Loop Test,"
Nucl. Phys. {\bf B452} (1995) 261; hep-th/9506126. }
\lref\mayr{P. Mayr, `` Mirror Symmetry, N=1 Superpotentials and
Tensionless Strings on Calabi-Yau Four-Folds,"  Nucl. Phys. {\bf
B494} (1997) 489; hep-th/9610162.}
\lref\wl{W. Lerche, ``Fayet-Iliopoulos Potentials from
Four-Folds," J. High Energy Phys. {\bf 11} (1997) 004;
hep-th/9709146.}
\lref\bb{ K. Becker and M. Becker, ``M-Theory on Eight-Manifolds,"
Nucl. Phys. {\bf B477} (1996) 155; hep-th/9605053.}
\lref\svw{S. Sethi, C. Vafa and E. Witten, ``Constraints on
Low-Dimensional String Compactifications," Nucl. Phys. {\bf B480}
(1996) 213; hep-th/9606122.}
\lref\afmn{I. Antoniadis, S. Ferrara, R. Minasian and K. S.
Narain, ``$R^4$ Couplings in M and Type II Theories," Nucl. Phys.
{\bf B507} (1997) 571; hep-th/9707013.}
\lref\fhmm{D. Freed, J.A. Harvey, R. Minasian and G. Moore,
``Gravitational Anomaly Cancellation for  $M$-Theory Fivebranes,''
Adv. Theor. Math. Phys. {\bf 2} (1998) 601; hep-th/9803205.}
\lref\syz{A. Strominger, S.T. Yau and E. Zaslow, ``Mirror Symmetry
is T-Duality,'' Nucl.Phys. {\bf B479} (1996) 243; hep-th/9606040.}
\lref\vafa{C. Vafa, ``Extending Mirror Conjecture to Calabi-Yau
with Bundles," hep-th/9804131.}
\lref\ch{C.G.~Callan and J.A.~Harvey, ``Anomalies And Fermion Zero
Modes On Strings And Domain Walls," Nucl. Phys. {\bf B250}(1985)
427.}
\lref\gvw{ S. Gukov, C. Vafa and E. Witten, ``CFT's From
Calabi-Yau Four-folds," hep-th/9906070.}
\lref\hpxx{P.S.~Howe and G.~Papadopoulos, ``Ultraviolet Behavior
Of Two-Dimensional Supersymmetric Nonlinear Sigma Models," Nucl.
Phys. {\bf B289} (1987) 264.}
\lref\hm{J.A.~Harvey and G.~Moore, ``Superpotentials and membrane
instantons,'' hep-th/9907026.}
%
%

\Title{\vbox{\baselineskip12pt \hbox{YCTP-P17-99}
\hbox{hep-th/9906190} }} {\vbox{\centerline{ M5-branes, Special
Lagrangian Submanifolds}
\medskip\centerline{
 and $\sigma$-models}
}}

\bigskip
\centerline{Ruben Minasian and Dimitrios Tsimpis}
\bigskip
\centerline{Department of Physics, Yale University}
\centerline{New Haven, CT 06520, USA}

\bigskip
\centerline{\bf Abstract}
\medskip

We study M-theory fivebranes wrapped on Special Lagrangian
submanifolds ($\S_n$) in Calabi-Yau three- and fourfolds. When the
M5 wraps a four-cycle, the resulting theory is a two-dimensional
domain wall embedded in three-dimensional bulk with four
supercharges. The theory on the wall is specified in terms of the
geometry of the CY manifold and the cycle $\S_4$. It is chiral and
anomalous, however the presence of a three-dimensional
gravitational Chern-Simons terms with a coefficient that jumps
when crossing the wall allows to cancel the anomaly by inflow.
K\"ahler manifolds of special type, where the potential depends
only on the real part of the complex coordinate, are shown to
emerge as the target spaces of two-dimensional $\s$-models when
the M5 is wrapped on $\S_3 \times S^1$, thus providing a physical
realization of some recent symplectic construction by Hitchin.

\Date{June, 1999}

%

\newsec{Introduction and summary}

Calibrated geometries \hl\ (see also \refs{\harv, \lm}) have been
studied in the context of mirror symmetry \refs{\syz, \dm} and
intersecting branes (see e.g. \fof\ for a review and references).
Our focus is on Special Lagrangian (SL) submanifolds in Calabi-Yau
three- and fourfolds,  and we mostly concentrate on the
deformation properties of these submanifolds.

We study a class of two-dimensional $\sigma$-models originating
from $M$-theory fivebranes wrapped on four-manifolds. Fivebranes
wrapped on holomorphic  divisors in a Calabi-Yau threefold were a
subject for recent investigation \refs{\msw, \mmt}. Here we turn
to the case of M5 wrapped on SL four-cycles, when $M$ theory is
compactified on a Calabi-Yau fourfold. While the amount of
supersymmetry can be figured out by general arguments \bbs, the
actual counting of the multiplets is somewhat involved. Such a
counting was done for a (very ample) divisor $\cp$ in a CY
threefold, in \msw. The geometry of the resulting $(0,4)$
$\s$-model was discussed in detail in \mmt. The understanding of
cases with lower supersymmetries, arising when M5 and D5 are
wrapped on a SL four-cycle in a Calabi-Yau four-fold, requires
some results from McLean's deformation theory \mcl. Recent
symplectic constructions by Hitchin also enter in a natural way
\refs{\hita, \hitb}. Indeed, there is a natural extension to the
symplectic case of many results on deformations of complex
submanifolds, and thus some of the results of \refs{\msw, \mmt}
(where the submanifold was taken to be a divisor in a CY
threefold) can be extended as well. Our study shows yet another
aspect of the intimate relation between calibrated geometries and
supersymmetry. Even though the main focus of our attention is on
SL submanifolds, other calibrations can be studied along similar
lines.

The wrapped fivebranes and the resulting $\sigma$-models are our
main concern, but it should be noted that the study of cycles is
relevant for D-brane moduli spaces, in particular instanton
effects in type $IIb$ ($F$-theory) compactifications when the
D3-branes are completely wrapped. The wrapping of D3-branes on
supersymmetric cycles in four-folds was considered in \bmoy, and
for the case of SL submanifolds such configurations were shown to
preserve half of supersymmetry.

In principle, there are few $\s$-models obtained by M5 wrapping SL
submanifolds in four-folds. The summary is presented in the table
($X_n$ denotes a (complex) $n$-fold of $SU(n)$ holonomy, while
$\Sigma_k$ denotes a SL cycle of real dimension $k$):

\bigskip
\begintable
Four-fold | SLS | Preserved Susy   | $\sigma$-model \elt $T^8$ |
$T^4$ | 1 | $(8,8)$ \elt $T^4 \times K3$ | $T^2 \times \Sigma_2$ |
${1/2}$ | $(4,4)$ \elt $K3 \times K3$| $\Sigma_2 \times
\Sigma_2^{\prime}$ | ${1/ 4}$ | $(2,2)$ \elt $T^2 \times X_3$ |
$S^1 \times \Sigma_3$ | ${1/4}$ |$(2,2)$ \elt $X_4$| $\Sigma_4$ |
${1/ 8}$ |$(0,2)$
\endtable
\bigskip
\noindent

In this note, we discuss only the multiplet structure and the
classical target spaces. In cases with lower supersymmetries that
are discussed here these will not be protected against quantum
corrections and there can be world-sheet renormalizations.

First, we will analyze the cases where the submanifolds are
complex concentrating on the fourfolds with $SU(4)$ holonomy\foot{
$K3 \times K3$ compactifications can be discussed as a special
case. }. Before going into details of the multiplet structure, we
can have a look at the field content after the reduction. As in
the case of \mmt\ the target space classically factorizes into two
sectors, the ``universal'' and the ``entropic'' (to borrow the
terminology from \mmt). The universal sector consists of two real
scalars, one coming from the coordinate parametrizing the position
of the M5 in three dimensional spacetime and one coming from the
component of the $\b$-field along the K\"{a}hler form (this will
be better explained in section 2). In the entropic sector the two
sources for the two-dimensional scalar fields are the self-dual
tensor field on the fivebrane worldvolume, and the deformations of
the cycle $\cc$ inside the Calabi-Yau. The position of the cycle
inside the fourfold is parametrized by the four (out of five)
scalars in the $(0,2)$ tensor multiplet on the fivebrane
worldvolume. After the reduction these will yield $d_{\cc}$ real
scalars on the world-sheet (the notation will be justified
shortly). The self-dual $\b$-field gives rise to $b_2^+$ and
$b_2^-$ right- and left-movers respectively (we have already
mentioned that one of the scalars coming from the $\b$ field
belongs to the ``universal'' sector). The excess of left-movers
tells us that the resulting $\s$-model is of heterotic type, and
there is a coupling to a gauge field. The numbers of right- and
left-moving fermions are given by the twisted Dirac index. A very
quick analysis of the multiplets reveals the connection between
supersymmetry and McLean's deformation theory \mcl: It is required
by supersymmetry (simply by boson-fermion matching) that the first
Betti number of the submanifold $b_1(\cc)= {\rm dim}H^1(\cc, \IR)$
be equal to the dimension of the moduli space of deformations of
$\cc$ inside $X$, $d_{\cc}$. Indeed it is known that the tangent
space of the local moduli space at $\cc$ of Special Lagrangian
submanifolds can be canonically identified with the space of
harmonic one-forms on $\cc$ \mcl. This identification is crucial
for the supersymmetry of the two-dimensional $\s$-model. The idea
of using supersymmetry transformations to give a physical
derivation of McLean's deformation theorems was explained to us by
J. Harvey and G. Moore (see also \hm).

Since the resulting two-dimensional theory is chiral, it can
suffer from anomalies. In section 3, we discuss the cancellation
mechanism. It turns out that the three dimensional bulk theory has
a gravitational Chern-Simons term with a discontinuity in its
coefficient. The variation of this term cancels the
two-dimensional anomaly by inflow.

In section 4, we will turn to M5 wrapped on  a real manifold $S^1
\times \Sigma_3$. By duality, this case will be related to the
study of moduli spaces of D-branes, and we will be able to
generalize some of the results here to D-branes wrapped on SL
cycles. Using the self-duality of the $\b$ field on the
worldvolume of M5 and the fact that $b_1(\S_3) = b_2(\S_3)$, it is
easy to see that the resulting number of zero-modes from the
tensor field is $b_1$. Just like for D-branes, these modes are
paired with $b_1$ zero-modes arising from deformations. However
supersymmetry predicts the stronger result that the target space
is K\"ahler.  In constructing the metric, we find many analogies
with the $c$-map construction of Cecotti, Ferrara and Girardello
\cfg\ and establish a {\it weak c-map.} The result is in agreement
with the one obtained for D-brane moduli spaces \hita. A special
care is needed for the treatment of the normal directions.

\newsec{Wrapped M5: Complex Special Lagrangian cycles}

Here we will perform the reduction to two dimensions of the 6D
supersymmetry. As explained in \mmt, it suffices  to consider the
reduction of the supersymmetry transformations of the tensor
multiplet from flat space. For this purpose we will need the
zero-mode expansion of the 6D fields in the fivebrane action.

\bigskip
\noindent {\it The bosonic zero-modes}

\medskip
\noindent The Kaluza-Klein ansatz for the chiral two-form $\b$ is:
\eqn\bkk{\b= \rho^a \omega_a +( \pi^I \omega_I +c.c.)+ uJ,}
where $J$ is the K\"{a}hler form on $\cc$, $\{ \omega_I, \, I=1,
\dots b_{20}(\cc) \}$ is a basis of $H^{2,0}(\cc)$ and $\{
\omega_a, \, a=1, \dots b_2^-(\cc) \}$ is a basis of
anti-self-dual $(1,1)$ forms on $\cc$. The scalars $\rho^a, \, u$
($\pi^I$) are real (complex). We therefore get $b_2^-$ left-moving
and $b_2^+$ right-moving real scalars. We have implicitly used the
fact that due to the K\"{a}hlerity of $\cc$ we have the
decomposition
\eqn\whtd{H^{2+}(\cc)=H^{2,0}(\cc) \oplus H^{0,2}(\cc) \oplus J;
\,\,\,\,\, H^{2}(\cc)=H^{2+}(\cc) \oplus H^{2-}(\cc)}

The five scalars of the 6D (2,0) tensor multiplet parametrize the
position of the fivebrane in transverse five-dimensional space.
When the fivebrane is wrapped, four of them (say $X^{6-9}$)
parametrize the position of the four-cycle ${\cal C}$ inside the
Calabi-Yau four-fold $X$ while the fifth $(X^{10})$ describes the
motion of the string in the three non-compact dimensions. The
massless modes arise from deformations of $W_2 \times \cc$
preserving supersymmetry. Let ${\cal M}$ be the space of
deformations of $\cc$. The tangent space to $\cm$ at $\cc$ is
\eqn\tstm{T_{\cc} \cm = H^0(\cc, \cn)}
where $\cn$ is the normal bundle of $\cc$ inside $X$. For $\cc$ a
special lagrangian submanifold the following equivalence holds
\mcl:
\eqn\eonttt{\cn \cong T^*\cc}
We will rely heavily on this relation in the following. Taking
into account that $H^0(\cc, T^*\cc) \cong H^{1,0}(\cc)$ we see
that \tstm\ implies
\eqn\tdom{dim_{\IC} \cm =b_{10}(\cc)={1 \over 2}b_1(\cc)}
Let us choose a complex basis $\{ \upsilon^m_{\ti}, \, m=1,2,
\,\, \ti =1, \dots b_{10}(\cc) \}$ of sections of $\cn$. Due to
the equivalence \eonttt\ we may identify
\eqn\eontttb{\upsilon^{1,2}_{\ti}= \omega^{1,2}_{\ti},}
where $\{ \omega_{\ti \bar{m}} dz^{\bar{m}}, \}$ is a basis of
$H^{0,1}(\cc)$ and raising/lowering of the $m$ index is possible
due to the existence of a metric on $T \cc$. We will expand to
first-order in $\varphi^{\tilde{I}}$
\eqn\sckkexp{X^6+iX^7=2\upsilon^1_{\ti}\varphi^{\ti}; \,\,\,
X^8+iX^9= -2\upsilon^2_{\ti}(\varphi^{\ti})^*}
where $\varphi^{\ti}, \,\, \ti=1, \dots {1 \over 2}b_1$ are
two-dimensional complex massless unconstrained bosons. The
(world-sheet) scalar $X^{10}$ accounts for one real boson.

Altogether the number of right-, left-moving real bosonic degrees
of freedom is
\eqn\nobdofs{N_L^B=b_1+b^-_2+1; \,\,\, N_R^B= b_1+b^+_2+1.}

\bigskip
\noindent {\it The fermionic zero-modes}
\medskip
\noindent

The fivebrane breaks the Lorentz invariance of $W_6$ down to
$Spin(1,1)_{W_2} \times Spin(4)_{\cc}$. The fermions of the tensor
multiplet transform in the $(+,{\bf 2}_+) \oplus (-,{\bf 2}_-)$.
Moreover the $Spin(5) \cong USp(2)$ ${\cal R}$-symmetry of the 6D
tensor multiplet is broken by the Calabi-Yau to $Spin(4)_{\cn}$,
where $SO(4)_{\cn}$ is the structure group of $\cn$. From the 2D
point of view there is a $Spin(4)_{\cc} \times Spin(4)_{\cn}$
${\cal R}$-symmetry and fermions in the $({\bf 2}_-,{\bf
2}_{\pm})$, $({\bf 2}_+,{\bf 2}_{\pm})$ give rise to left-,
right-movers on $W_2$ respectively. Due to \eonttt\ the fermions
can be thought of as bispinors on $\cc$ and are in one-to-one
correspondence with $(p,q)$ forms on $\cc$: Since $\cc$ is
K\"{a}hler, one can construct two sets of gamma matrices $\{
\gamma^m, \gamma^{\bar{n}} \}=\{ \tilde{\gamma}^m,
\tilde{\gamma}^{\bar{n}} \}= g^{m\bar{n}}, \,\, [ \gamma,
\tilde{\gamma} ]=0$. We can take
$\tilde{\gamma}^m=i\rho^{(\cc)}\gamma^m$, where $\rho^{(\cc)}$ is
the chirality matrix on $\cc$. We will regard $\gamma^m,
\tilde{\gamma}^{\bar{m}}$ as creation operators and let us denote
the ``Fock vacuum'' by $\vert 0 \rangle$. Let $\psi^R$ ($\psi^L$)
be a bispinor in the $({\bf 2}_+,{\bf 2}_+) \oplus ({\bf 2}_+,{\bf
2}_-)$ ($({\bf 2}_-,{\bf 2}_+) \oplus ({\bf 2}_-,{\bf 2}_-)$) of
$Spin(4)_{\cc} \times Spin(4)_{\cn}$. One has
\eqn\scfex{\eqalign{
 \psi^R&\sim (\Omega^{(0,0)}+ \Omega^{(0,1)}_{\bar{m}}
\tilde{\gamma}^{\bar{m}}+ \Omega^{(2,0)}_{mn} \gamma^{mn}
+\Omega^{(0,2)}_{\bar{mn}}\tilde{ \gamma}^{\bar{mn}}+
\Omega^{(2,1)}_{mn\bar{k}}\gamma^{mn} \tilde{\gamma}^{\bar{k}}+
\Omega^{(2,2)}_{mn\bar{kl}} \gamma^{mn}\tilde{ \gamma}^{\bar{kl}})
\vert 0 \rangle \cr \psi^L&\sim (\Omega^{(1,0)}_{m} \gamma^{m}+
\Omega^{(1,1)}_{m\bar{n}} \gamma^m
\tilde{\gamma}^{\bar{n}}+\Omega^{(1,2)}_{m\bar{nk}}\gamma^{m}
\tilde{\gamma}^{\bar{nk}} ) \vert 0 \rangle \cr} }
where $\Omega^{(p,q)}$ is a $(p,q)$ form on $\cc$. To see this
note that because $\cc$ is K\"{a}hler the positive and negative
spin bundles on $\cc$ (due to \eonttt\ a similar result holds for
$\cn$) decompose as:
\eqn\spindecomposition{S^+(T\cc)\otimes K^{1/2} \cong
\Omega^{0,0}\oplus
 \Omega^{2,0}; \,\,\,\,\,\, S^-(T\cc)\otimes K^{-1/2}
 \cong \Omega^{1,0}}
and we can expand the fermionic zeromodes in terms of forms on
$\cc$. Also note that $\Omega^{(0,1)}_{\bar{m}}
\tilde{\gamma}^{\bar{m}} \vert 0 \rangle$ ($\Omega^{(1,0)}_{m}
\gamma^{m} \vert 0 \rangle$) transforms in the ${\bf 2}_+$ (${\bf
2}_-$) of $Spin(4)_{\cn}$ ($Spin(4)_{\cc}$).
Taking into account \whtd\ and the isomorphism $H^{p,q}(\cc)\cong
H^{2-p,2-q}(\cc)$ we see that from the 2D point of view the number
of (real) left- and right-moving fermionic degrees of freedom is
\eqn\nodofs{N_L^F=b_1+b^-_2+1; \,\,\, N_R^F= b_1+b^+_2+1.}
In order to reduce the supersymmetry, we need to be slightly more
explicit than in \scfex. Let us choose our ten-dimensional
matrices to locally decompose as
\eqn\sxmgamma{\eqalign{\Gamma^{0,1} &= \gamma^{0,1}
\otimes\rho^{(\cc)} \otimes\rho^{(\cn)}  \cr \Gamma^{2,3,4,5} &=
{\I1}_2 \otimes \gamma^{2,3,4,5} \otimes \I1^{(\cn)} \cr
\Gamma^{6,7,8,9} &= {\I1}_2 \otimes \rho^{(\cc)} \otimes
\gamma^{6,7,8,9} \cr}}
where
\eqn\gmzroonetfv{\eqalign{\gamma^0=i\sigma^2&;\,\,\,\,\,
\gamma^1=\sigma^1 \cr \gamma^{2,3,4} = \pmatrix{ 0 &
\sigma^{1,2,3} \cr \sigma^{1,2,3} & 0 \cr }&; \,\,\,\,\,
\gamma^{5} = \pmatrix{ 0 & i \, \I1_2 \cr -i \, \I1_2 & 0 \cr }
\cr}}
and $\gamma^i= \gamma^{i+4}, \, i=2, \dots 5$.

We locally decompose the covariantly constant spinor of the
Calabi-Yau as $\xi^{(8)}=\xi^{(\cn)} \otimes \xi$ and we take
$\xi^{(8)}, \, \xi$ to be of positive chirality. If we write
$\xi^{(8)}$ as $\xi_{(i)}$ with $i=1, \dots 4$ the symplectic
index, the statement about the chirality corresponds according to
our conventions to setting $\xi_{i=3}= \xi_{i=4} =0$. The fermions
are expanded in terms of 2D right- and left-movers as
\eqn\tzspin{\psi^{(6)}_i = \psi^I_{i-} \otimes \Delta^I_{(i)} +
\psi^{\ti}_{j-} \otimes \Delta^{\ti}_{(ij)} + \psi^0_{i-} \otimes
\xi_{(i)} + {\rm left } \, {\rm movers}, }
where
\eqn\defdelta{ \Delta_{I(i)} = \cases{ \omega^{I\bar{mn}}
\gamma^{\bar{mn}} \xi  & for $i=1$ \cr \omega_{Imn} \gamma^{mn}
\xi^* & for $i=2$
 \cr 0 & for $i=3,4$}
; \,\,\,\,\,  \Delta_{\ti(ij)}= (\omega_{\ti \bar{m}}
\gamma^{\bar{m}}+ \omega_{\ti m} \gamma^{m})_{ij} \xi_{(j)} }
and
\eqn\defxi{ \xi_{(i)} =\cases{ \xi  & for $i=1$ \cr \xi^* & for
$i=2$ \cr 0  & for $i=3,4$}}
The right-moving zero modes $\psi^I_{1,2}, \, \psi^{\ti}_{1,2}, \,
\psi^{0}_{1,2}$ are complex antichiral 2D fermions. Only half of
those are independent since the 6D symplectic reality condition
implies $\psi^x_{1}=i(\psi^x_{2})^*$, where $x=0, \, \ti, \,$ or
$I$. In the following we set $\psi^x:=\psi^x_1$.

\subsec{The 2d supersymmetry}

Substituting the expansions \bkk, \sckkexp, \tzspin\ into the
supersymmetry transformations of the tensor multiplet we get
\eqn\tdsusy{\eqalign{\delta \varphi^{\ti}&=- \varepsilon^*_+
\psi^{\ti}_-; \,\,\,\,\, \delta \psi^{\ti}_-=
\partial_-\varphi^{\ti} \varepsilon_+ \cr \delta \pi^{I}&=-
\varepsilon^*_+ \psi^{I}_-; \,\,\,\,\, \delta \psi^{I}_-=
\partial_-\pi^{I} \varepsilon_+ \cr \delta \rho^a&=0;
\,\,\,\,\,\,\,\,\,\,\,\,\,\,\,\,\,\,\,\,\,\, \delta \psi^{a}_+ =
0\cr}}
where $\psi^{a}_+$ are the left-moving fermions and the
supersymmetry parameter $\varepsilon_+$ is a chiral complex 2D
spinor. These are the supersymmetry transformations, written in
complex notation, of the 2D fields of a $(0,2)$ $\sigma$-model. In
order to compare with the standard $(0,1)$-superspace description
\hw\ and read off the complex structure of the target space of the
$\sigma$-model, we pass to a real basis:
$\pi^I=\phi^{1I}+i\phi^{2I}; \,\,
\varphi^{\ti}=\phi^{1\ti}+i\phi^{2\ti}$. Similarly for the
fermions: $\psi^I_-=\chi^{1I}_-+i\chi^{2I}_-; \,\,
\psi^{\ti}_-=\chi^{1\ti}_-+i\chi^{2\ti}_-; \,\, \varepsilon_+=
\zeta_+ +i \eta_+$. The susy transformations become
\eqn\rsus{\eqalign{\delta \phi^{ix}&= (\zeta_+ \delta^{ix}{}_{jy}+
\eta_+ J^{ix}{}_{jy})\chi^{jy}_- \cr
 \delta \chi_-^{ix}&= (\zeta_+ \delta^{ix}{}_{jy}- \eta_+
J^{ix}{}_{jy}) \pam \phi^{jy};  \,\,\,\,\, i=1,2; \,\,\, x,y=I \,
{\rm or} \,
\ti
\cr}}
where $J^{ix}{}_{jy}:=i[\sigma^2]^i{}_j \delta^x{}_y$ satisfies
$J^2=-\I1$. Moreover the transformations \rsus\ and the fact that
$(0,2)$ supersymmetry is unbroken implies that $J$ has to be
covariantly constant, giving a K\"{a}hler target space.

The transformation of $\psi^0$ can be joined with the
transformations for $u$ and $X^{10}$ to give
\eqn\untr{\delta \psi^0_-= \pam U \varepsilon_+; \,\,\,\, \delta
U= -\varepsilon_+^* \psi^0}
where $U:=X^{10}+iu$. This ``universal'' factor, consisting of the
fields $\{u, X^{10}, \psi^0_- \}$, possesses separately $(0,2)$
supersymmetry.

\subsec{The structure of the $\s$-model}

As we have seen, the spectrum on the two-dimensional worldvolume
is determined by three numbers $b_2^+$, $b_2^-$ and $b_1$.
Similarly to \msw, one can express $\s(\cc) = b_2^+ - b_2^-$ and
$\chi(\cc) = 2 + b_2^+ + b_2^- - 2 b_1$ in terms of Calabi-Yau
quantities
\eqn\eulpsig{\chi(\cc) = {\eta}^2; \,\,\,\,\, \s(\cc) = - {1 \over
3} c_2{\eta} }
where
\eqn\defsd{c_2 {\eta}:= \int_X c_2(X) \wedge {\eta};
 \,\,\,\,\, {\eta}^2:=\int_X {\eta}
\wedge {\eta}}
and ${\eta}$ is the Poincar\'e dual to the cycle $\cc$. It is an
element of $H^4(X, \IZ)$ \foot{Strictly speaking, ${\eta}$
 is an element of the
cohomology with compact support $H_{cpct}^4(X, \IZ)$ which in
general for noncompact $X$ forms a sublattice of $H^4(X, \IZ)$.
For a detailed discussion in a similar context see \gvw.} and
restricts to  ${\eta}=c_2(\cn)= c_2(\cc)$ on $\cc$ \refs{\grha,
\hub}, where for the last equality we used \eonttt. Equation
\eonttt\ and the exact sequence
\eqn\exctsq{0 \rightarrow T\cc \rightarrow TX \rightarrow \cn
\rightarrow 0}
were also used in order to derive \eulpsig.

Since $\cc$ is holomorphically embedded in $X$, ${\eta}$ is of
type $(2,2)$. The latter requirement (the existence of a
holomorphic four cycle) puts a restriction to the complex
structure of $X$. Note also that ${\eta}$ is primitive\foot{ This
can be seen roughly as follows: We locally parametrize X by
$(l_a,n^a), \, a=1, \dots 4$ so that $\cc$ is given by $n^a=0$.
Then ${\eta}$ is schematically given by $\d (n) dn^1 \wedge \dots
dn^4$. Since $\cc$ is SL we can locally write the K\"{a}hler class
as $\omega =dl_a \wedge dn^a$. We thus see that $\omega \wedge
{\eta} =0$}. In the next section we will explain the relation
between ${\eta}$ and the cohomology class of the four-form field
strength $G_4$ of eleven dimensional supergravity. The conditions
that $G_4$ is integral, $(2,2)$ and primitive are precisely the
requirements for compactification of M-theory on a manifold of
eight (real) dimensions preserving four supercharges in three
dimensional Minkowski spacetime \bb.

Finally putting everything together, we see that the total number
of $(0,2)$ multiplets is $b_1 + b_2^+ + 1,$ while the rank of the
vector bundle over the target space is
\eqn\rank{{\rm rank}V = {1 \over 2} (b_1 + b_2^- +1) + \vert
\s(\cc) \vert .}
Note that the model is chiral even when the numbers of left- and
right-moving modes coming from the $\b$-field are the same
($\s(\cc) = 0$). The standard formulation with $(0,1)$
supersymmetry \hw\ couples the gauge field to the fermionic
current. The latter can be bosonized as in \mmt. Here the
left-moving (bosonized) current comes from both the bosonization
of the left moving fermions and from the $\vert \sigma (\cc)
\vert$ left-moving bosons.

As in \mmt, in reducing the fivebrane action to the worldsheet of
the $\sigma$-model, both the $\b$-field and the gauge connection
of the vector bundle will appear flat at least to this
approximation. Repeating the analysis of \mmt\ to extract
information for the metric, will involve variations of Hodge
structures of weights one and two. However the classical geometry
of the model is not protected by supersymmetry from quantum
corrections \hpxx\ and it is not clear to what extent such an
analysis should be trusted.

When the fourfold is of the form $K3 \times K3$, the corresponding
SL submanifold is still complex, and  the resulting
two-dimensional theory can be discussed as a special case of the
more general model presented above. Now one has to bear in mind
that the fermions on the CY are no longer chiral.  As before, for
each  cycle $\S_2^i$, $i=1,2$ we can take ${\eta}_i = c_1(N^i) =
-c_1(\S^i)$ and have $b_1^i = 2D_i +2$ where $D_i ={1 \over 2}
\int_{X_i} {\eta}_i^2.$ Then for $\S_2^1 \times \S_2^2$,
\eqn\kbet{\eqalign{b_1 &= 4 + 2D_1 + 2D_2 \cr b_2 &= 2+ (2D_1+2)
(2D_2 +2) \cr}}
and $b_2^+ = b_2^-$. We can see that for the left-movers the gauge
bundle can be identified with the tangent bundle, and the
resulting model has $(2,2)$ supersymmetry in accordance with the
expectations from the general supersymmetry analysis.

\newsec{Inflow in 3 dimensions}

Following \hmm, we would like to present an alternative way of
counting of the zero-modes on the string.  This counting is based
on using the  fivebrane anomalies. Here as well we can reduce the
fivebrane anomaly given by \wittfive\
\eqn\anoma{I_8 (TW, \cn) = {1 \over 48}\Biggl({1 \over 4} (p_1(TW)
- p_1(\cn))^2 + p_2(\cn) - p_2(TW)\Biggr)}
by using $TW_6 \vert_{\cc} = T\cc \times TW_2$ and $\cn =\cn (\cc
\hookrightarrow X)$. On the other hand we know that the total
two-dimensional anomaly (note that there is no anomaly of the
normal bundle anymore) is given simply by the difference of
central charges of left- and right-movers, $I_{strng} = (c_L -c_R)
p_1(TW_2)/24.$ Comparing with the reduction of the fivebrane
anomaly \anoma, we recover
\eqn\diff{c_L - c_R = \half c_2 {\eta} = -{3 \over 2} \s(\cc)}
in agreement with \nobdofs, \nodofs\ and \eulpsig.

We now turn to the anomaly cancellation mechanism and find some
new twists here. In particular, the discussion of anomaly
cancellation is closely related to the jump in the value of a
certain cohomology class (to be identified momentarily)  recently
discussed in \gvw. Since in eleven dimensions a complete
cancellation of the anomalies on the fivebrane worldvolume occurs
\fhmm,  we expect the same to happen after the
compactification/wrapping. Note that in three dimensions, the
string is a domain-wall type of object and thus it is magnetically
charged with respect to the zero-form field strength whose value
jumps when crossing the wall. The source equation for such a
string can be written as
\eqn\source{d \Lambda = \d(W_2 \hookrightarrow M_3)}
where $\d(W_2 \hookrightarrow M_3)$ is the Poincar\'e dual to the
string worldvolume. With some abuse of notation we will call
$\Lambda$ a ``cosmological constant"; it takes different values on
the different sides of wall\foot{One should not confuse this with
the case where there is an actual cosmological constant in the
uncompactified dimensions, corresponding to the geometry of
M-theory on $AdS_3 \times X_4$ \gvw. There the cosmological
constant turns out to be proportional to the projection of the
four-form field strength of 11D supergravity to the $(4,0)$ part
of the cohomology of $X_4$.  }.

Equation \source\ can be seen as simply the reduction of the
fivebrane source equation. Somewhat schematically, in this
background the fivebrane source equation becomes
\eqn\fdse{d G_4 =2 \pi \delta(x)dx \wedge {\eta},}
where $x$ is a coordinate along the direction transverse to $W_2$
in $M_3$ so that $\delta(x)dx$ represents the Poincar\'e dual of
$W_2 \hookrightarrow M_3$. Similarly ${\eta}$ is the Poincar\'e
dual of $\cc$ inside $X$ and can be represented as a four-form
with compact support $\d(\cc_4 \hookrightarrow X)$. We see that
the cohomology class of $G_4/2 \pi$ jumps by $\eta$ when crossing
the wall at $x=0$ in $M_3$.

As seen in \hmm, when $G_4$ has non-zero fluxes through a
four-dimensional surface, the term in the effective action \dlm,
\eqn\dent{\Delta_{11} = G_4 \wedge X_7^{(0)}(TM),}
where
\eqn\anpl{dX_7^{(0)}(TM) = -{1 \over 48} ({1\over 4} p_1^2(TM) -
p_2(TM))),}
can give rise to lower-dimensional Chern-Simons terms. Indeed
there is a flux of the four-form field strength through a
four-cycle and the reduction yields a three-dimensional
gravitational Chern-Simons term:
\eqn\redu{ \sim{1 \over 48} \int_X {\eta} \wedge p_1(T\cc) \cdot
p_1^{(0)}(TM) \sim {3 \over 2} \s(\cc) {p_1^{(0)}(TM)\over 24}, }
where
\eqn\lkju{dp_1^{(0)}(TM)= p_1(TM)}
As we have already seen, the string anomaly is $\sim {3 \over
2}\s(\cc)$ as well. To complete the discussion we need to examine
the flux quantization.

The compactification of $\Delta_{11}$ and the Chern-Simons
coupling of the eleven-dimensional supergravity on fourfolds leads
to a tadpole in three dimensions \refs{\svw, \wof} proportional to
$(\chi/24 - \int_X G_4^2/(8\pi^2))$ where $\chi$ is the Euler
number of the fourfold $X$. The coefficient of the
three-dimensional gravitational Chern-Simons term is expected to
be the same by supersymmetry, and we argue that the coupling is of
the form:
\eqn\rrr{\Delta_3 = {3 \over 2} \, \Lambda \,
\s(\cc)p_1^{(0)}(TM)}
which taking \source\ into account cancels by inflow the string
anomaly. It would be interesting to check the three-dimensional
supersymmetry of these terms directly. Note that a similar
structure of a gravitational Chern-Simons term with discontinuity
in the coefficient is also expected from the general discussion of
inflow on domain walls \ch.

\newsec{$(2,2)$ $\s$-models, D-brane moduli spaces and
the weak $c$-map}

By wrapping the $M5$ on $\S_3$ in a CY threefold $X_3$, we arrive
at a three dimensional theory with four supercharges ($N=2$).
Further wrapping on $S^1$ gives a $(2,2)$ two-dimensional model
that can be directly obtained from a  D4 wrapped on $\S_3$. As we
will see the results easily generalize to all D-branes.

Because of the self-duality of the $\b$-field on the worldvolume
of M5 and due to the fact that $b_1(\S_3) = b_2(\S_3)$, we can get
either a three-dimensional theory with $b_1$ vectors or a theory
with $b_1$ scalars, which are dual to each other in three
dimensions. In addition there are $d_{\cc} = b_1$ zero modes,
coming from deformations of the SL submanifold inside the CY
threefold, and a ``universal'' sector consisting of two real
bosons (and their susy partners completing a $(2,2)$ multiplet)
coming from the coordinates parametrizing the position of $W_2
\times S^1$ inside $\IR^{1,2} \times T^2$. When $b_1(\S_3) = 0$,
the cycle is rigid, and the only surviving degrees of freedom in
two dimensions are the two real scalars of the universal sector.
The resulting CFT with ${\hat c} =3$ can be written down
explicitly \foot{It is probably interesting to note that this
theory corresponds to a SL three-manifold with vanishing first and
second Betti numbers. Such manifolds can be constructed by
quotiening $S^3$ by discrete groups, and they have non-trivial
first homotopy groups. It is an open conjecture that any closed
simply-connected (and therefore with vanishing $H_1$)
three-manifold is homeomorphic to $S^3$ (Poincar\'e conjecture).
}. In direct analogy with the construction of \cfg, we see that
the dualization of vectors leads to a doubling of the scalar
``entropic'' coordinates. It is clear that this doubling should
work in such a fashion to ensure the 3d $N=2$ supersymmetry. We
see that indeed the (real) moduli space of the deformations $\CM$
is such that $T\CM$ is a K\"ahler manifold. A K\"{a}hler metric on
this manifold was constructed by Hitchin in \hita. We can now see
a physical construction.

Fix a point $\{ \phi^i \}$ in the moduli space ${\cal M}$. At this
point the tangent space is
\eqn\tstm{T_{\phi}\cm =H^1(\cc, \IR) \oplus H^1(\cc, \IR) \cong
H^1(\cc, \IC) }
Let $(l_a, \, n^a(\phi))$ be a parametrization of $X$ such that
${\cal C}_{\phi} \hookrightarrow X$ is given by $n^a(\phi)=0$ and
$\{ l_a \}$ are coordinates along $\cc_{\phi}$. The normal
directions $X^a$ are KK expanded in terms of the basis $\{
\upsilon_i^a:= \partial n^a / \partial \phi^i \}$ of sections of
the normal bundle of ${\cc}_{\phi} \hookrightarrow X$ as
\eqn\xkkexp{\partial_{\mu} X^a = \u^a_i \partial_{\mu}  \phi^i,}
where $\partial_{\mu} :=\partial / \partial \sigma^{\mu} \,$
$\partial_{i} :=\partial / \partial \phi^{i} \,$, and
$\sigma^{\mu}; \, \mu=0,1,2$ are coordinates on $W_3$. The
$\phi^i$'s are real.

Using the special lagrangian property of ${\cal C}$ one can show
\hita\ that $\{ \omega_i; \, i=1, \dots b_1({\cal C}) \}$, where
$\omega_i := dl_a \upsilon_i^a$, is a basis of $H^1({\cal
C}_{\phi}, \IR)$. The hermitian metric on $H^1(\cc_{\phi}, \IR)$
defines a hermitian metric on $T \cm$:
\eqn\mtrcn{D_{ij}:= \int_{\cc} \omega_i \wedge \ast \omega_j =
\int_{\cc} dV_3 \sqrt{g} g_{ab} \upsilon_i^a \upsilon_j^b, }
where $g_{ab}$ is the metric on $\cc$. Let us define ``inertial''
bases $\{ \alpha_I \}$ of $H^1(\cc, \IR)$ and $\{ \b_I \}$ of
$H^2(\cc, \IR)$ by
\eqn\inerbs{  \omega_i = \Lambda_i{}^I \alpha_I ; \,\,\,\, \ast
\omega_i = M_i{}^I \b_I}
And let $\{ A_I \}$, $\{ B_I \}$ be the dual bases so that
\eqn\dotrdbs{\Lambda_i{}^I = \int_{A_I}\omega_i; \,\,\,\, M_i{}^I
= \int_{B_I} \ast \omega_i }
Hence the metric can be rewritten as
\eqn\vlbne{D_{ij}=\delta_{IJ} \Lambda_i{}^I M_j{}^J }
Since $\omega_i d\phi^i \,$, $\ast \omega_i d\phi^i $ are closed
as forms on $\cm$ \hita, we have
\eqn\cnsonmt{\partial_{[i} \Lambda_{j]}{}^I=0; \,\,\,\,
\partial_{[i} M_{j]}{}^I=0  }
and therefore we can locally write
\eqn\khlpt{D_{ij}=\partial_i \partial_j K(\phi)  }
for some scalar function $K(\phi)$.

Let us now use the above results to reduce the 6D theory of the M5
to $W_3$. Like before (see \mmt\ for detailed discussion of the
reduction and for references) it turns out that it suffices to
work with unconstrained $\b$-field in the action and impose the
self-duality condition as an extra constraint. The KK ansatz for
the tensor field is
\eqn\tenexp{\b_2 = \omega_{ia} A_{\mu}^i d\sigma^{\mu} \wedge dl^a
 + {1 \over 2} (\ast \omega_i)_{ab} \l^i dl^a \wedge dl^b
+ \beta_{\mu \nu}  d\sigma^{\mu} \wedge d\sigma^{\nu}}
Suppressing internal derivatives first and then imposing the
constraint $d\beta_2= \ast d\beta_2$ we get
\eqn\dbtaex{\eqalign{d\beta_2 &= d\sigma^{\mu} \wedge
\partial_{\mu} \beta_2 \cr &={1 \over 2} \epsilon_{\mu
\nu}{}^{\rho} ({\nabla}_{\rho} \l)^i \omega_{ia} d\sigma^{\mu}
\wedge d\sigma^{\nu} \wedge dl^a +{1 \over 2} ({\nabla}_{\mu}
\l)^i (\ast \omega_i)_{ab} d\sigma^{\mu} \wedge dl^a \wedge dl^b
\cr}}
where
\eqn\nbldf{ ({\nabla}_{\mu})^i{}_j:= \delta^i{}_j \partial_{\mu}
+\partial_{\mu} \phi^k \Gamma^i_{jk} }
and
\eqn\chrsmb{\Gamma^i_{jk}:= \Lambda^i{}_I \partial_j M_k{}^I}
Conceptually this is the connection, with respect to the basis $\{
\ast \omega_i\}$, on a fibre bundle with fibre $H^2(\cc_{\phi},
\IR)$ over the point $\phi^i \in \cm$. The action is \foot{As we
already noted above (4.9) the $\b$-field should be thought of as
being unconstrained, otherwise the first term on the rhs of (4.13)
vanishes identically. The self-duality is imposed as a constraint
on the equations of motion deriving from (4.13).}
\eqn\actunrd{S=\int_{W_6} d\beta_2 \wedge \ast d\beta_2+
\int_{W_3} d^3\sigma \int_{\cc} dV_3 \sqrt{g} g_{ab}
\partial_{\mu} X^a \partial^{\mu} X^b   }
Plugging \xkkexp, \dbtaex\ above \foot{ In \mmt\ it was shown that
the reduction of the fivebrane action in arbitrary curved
background is recovered (at least to leading order) by first
reducing the action in flat background and then covariantizing. }
and taking \mtrcn\ into account we get
\eqn\facti{S=\int_{W_3} d^3\sigma D_{ij}(\p)( \pa_{\m} \p^i
\pa^{\m}\p^j + ({\nabla}_{\m} \l)^i (\nabla^{\m}\l)^j )  }
From \cnsonmt\ we see that $\Lambda_i{}^I d{\phi}^i$ is a closed
1-form on $\cm$ and therefore we can introduce the coordinate
$u^I$ such that
\eqn\ucor{du^I:= \Lambda_i{}^I d{\phi}^i }
We moreover define
\eqn\ydf{y^I:=  M_i{}^I \l^i}
Using \vlbne, \ucor, \ydf\ it is easy to see that
\eqn\cpli{z^I:=u^I+i y^I}
defines an almost complex structure on $T\cm$, which can be shown
to be integrable, reproducing the construction of \hita\ in a
physical context.

Eq. \ucor\ defines a set of coordinates which can be shown to be
equivalent to the special coordinates of \vafa. Indeed we have
\eqn\eqtv{du^I= \int_{A_I} \Omega }
where $\Omega:= \omega_j \wedge d\phi^j$. We can ``normalize'' our
coordinates $\phi^i$ so that at $\phi^i=0$, $\omega_i= \a_I$ and
therefore $d\phi^i=du^I$ (see \inerbs, \ucor). The supersymmetry
transformations of section 2 are worked out at $\phi^i=0$. Note
that the resulting space $T\cm$ is K\"ahler, as follows from
\khlpt, and has the special property that the potential depends
only on the real part of the complex coordinates. Due to this very
restricted nature of the target space one may even expect some
non-renormalization theorems in spite of having little
supersymmetry. It will be interesting to investigate this question
in more detail.

To summarize, the target space of the $\s$-model obtained by
wrapping M5 on $\S_3 \times S^1$ is given by the moduli space of a
D3-brane wrapped completely on such a cycle. Due to the presence
of the $S^1$ and the equivalence of 3D vectors and scalars
($T$-duality), it is clear that wrapping a D5 would produce a
similar target space.

We conclude with two remarks on four-cycles. The case of a D5
wrapped on a cycle $\S_4$ in a fourfold is somewhat puzzling: it
would produce a space-filling two-dimensional theory that by
general arguments \bbs\ is expected to have very low
supersymmetry. From the other side as seen from \hitb, in this
case the cycle is complex and K\"ahler and the cotangent bundle
construction yields a hyper-K\"ahler target space indicating at
least four supercharges, and thus a supersymmetry enhancement.

Considering wrapped D-branes on SL leads to an identification
\syz\ of the complexified space $H^1(\cc)$ and    the moduli
spaces of stable bundles $V$  on the mirror manifold $Y$, $H^1
(End V, Y).$ It was conjectured in \vafa\ that this relation is
expected to generalize to $H^k(\cc) = H^k (End V, Y),$ for all
$k$. A higher-rank space, $H^2(\cc)$, appears for the first time
at $dim_{\IR}\cc=4$ and is relevant only for wrapped M5-branes
(while D-branes don't ``see" this space). It would be interesting
to understand if the M5-brane can play any role in testing the
extended mirror conjecture.

\bigskip
\centerline{\bf Acknowledgments}\nobreak
\bigskip

We are grateful to Gregory Moore for many important discussions
and for corrections to a preliminary version of the manuscript.
Conversations with Jeffrey Harvey, David Morrison and Erich
Poppitz are also gratefully acknowledged. This work is supported
by DOE grant DE-FG02-92ER40704. RM would like to thank the
hospitality of Ecole Polytechnique and LTPHE, Paris VI-VII during
the course of the work.

\listrefs
\end